\documentclass{PoS}

\title{Targeting Earth: CRPropa learns to aim}

\ShortTitle{Targeting Earth: CRPropa learns to aim}

\author{Jens Jasche$^1$, \speaker{Arjen van Vliet}$^{\;2,4}$, J\"org P. Rachen$^{3,4}$\\
        $^{1\,}$The Oskar Klein Centre, Department of Physics, Stockholm University,\\ AlbaNova University Centre, SE 106 91 Stockholm, Sweden\\
        $^{2\,}$Deutsches Elektronen-Synchrotron (DESY), Platanenallee 6, 15738 Zeuthen, Germany \\
        $^{3\,}$Astrophysical Institute, Vrije Universiteit Brussel, Pleinlaan 2, 1050 Elsene, Belgium \\
        $^{4\,}$Department of Astrophysics, Institute for Mathematics, Astrophysics and Particle Physics,\\ Radboud University, Postbox 9010, 6500 GL Nijmegen, the Netherlands\\
        E-mail: \email{arjen.van.vliet@desy.de}
        }

\abstract{Realistic predictions for the arrival directions of ultra-high-energy cosmic rays require extensive simulations of UHECR propagation through 3D space, potentially even including cosmological evolution and timing effects. Such 3D or 4D simulations of cosmic-ray propagation suffer from the fact that a relatively small target -- the observer sphere -- needs to be hit. If particles are ejected in any direction from the source according to the source emission geometry, such simulations are tremendously inefficient. We present here a targeting mechanism which finds an optimal emission geometry to maximize the number of hits while remaining unbiased in the arrival-direction distribution. This can lead to speedups by many of orders of magnitude, depending on the simulation setup. We present the basic mathematics to produce unbiased results from targeted simulations, demonstrate its effectiveness with the simulation package CRPropa\,3 for various propagation scenarios, and discuss prospects to include this mechanism as a standard part of CRPropa in the future.
}

\FullConference{36th International Cosmic Ray Conference -ICRC2019-\\
		July 24th - August 1st, 2019\\
		Madison, WI, U.S.A.}

\begin{document}

\section{Introduction}
\label{sec:intro}

\noindent The identification of the sources of ultra-high energy cosmic rays is still an open issue, and in fact a tricky business. The excellent information from the two large experiments, the Pierre Auger Observatory in the South, and the Telescope Array in the North, on spectrum~{\cite{Fenu:2017hlc, Matthews:2017hlc}} and chemical composition~{\cite{Unger:2017fhr, Abbasi:2018nun}} hardly constrain source scenarios, as these properties are difficult to predict from first principles and are usually chosen in any scenario suitably to fit the data. A new hope to eventually identify sources has been risen by the tentative claim of intermediate-scale anisotropies of the arrival directions of UHECR above a few tens of EeV~{\cite{Matthews:2017hlc, Aab:2018chp}}. If they can be further corroborated, one may compare them with sky-distributions of putative sources.

Given the complexity of the physics of cosmic-ray acceleration and propagation, however, it can hardly be assumed that na\"ive comparisons with astronomical catalogs will be conclusive. What needs to be done is to develop a parameterized source model attached to known individual astronomical objects (an example for radio galaxies is presented in these proceedings by Rachen \& Eichmann, \pos{PoS(ICRC2019)396}), as well as a realistic setup for extragalactic and Galactic magnetic fields including the uncertainties in their modeling. The propagation of UHECR in such a setup has then to be probed in extensive simulations, following the individual trajectories of a very large number of particles. Only the results of such simulations, when compared with real arrival direction distributions, will allow to conclusively constrain UHECR source scenarios. The currently most advanced tool to perform such simulations is CRPropa~\cite{2016JCAP...05..038A}, which has all processes relevant for UHECR propagation implemented and can apply them to calculate realistic 3D trajectories, potentially even considering cosmological evolution during long propagation path (then called a 4D simulation). 

There is a fundamental problem with these kind of simulations, though. Sources of UHECR are almost certainly extragalactic~{\cite{Aab:2017tyv}}, and may be spread over a volume of at least 300\,Mpc radius. Neglecting effects of our Galaxy for the time being, we may then consider a putative detector -- called an observer sphere -- of the order of the size of our Galaxy, roughly 30\,kpc, placed in the center of this volume.\footnote{
The reason for this common choice is not that propagation effects inside the Galaxy is ignored, but that a different method is used in CRPropa to describe them.
We will use the terms ``observer sphere'' and ``detector'' synonymous throughout this paper. 
}
The task to hit this observer sphere is comparable to hit the inner bull of a dartboard from 12 meters distance -- with the darts blown around by strong whirlwinds (i.e., the magnetic fields). The worst part of this dart game, however, is that the players' eyes are blindfolded and they do not know where the board hangs. 

While the large distance and the whirling are given by nature and we have to deal with them, the blindfolding is actually an unnecessary complication of the game. If we allow ``the players'' (i.e., our cosmic-ray simulation) to ``aim'', we can significantly increase the number of ``hits'', i.e., simulation candidates which can be used for astrophysical analysis, while at the same time avoiding any bias in the obtained results as long as we are able to correct for the aiming in a mathematically proper way. In this paper, we introduce the mathematical foundations of such a targeting mechanism, demonstrate its efficiency in a prototype implementation with CRPropa\,3, and briefly discuss how it may be implemented as a fixed part of future CRPropa versions.

\section{The probability distribution of event counts}
\label{sec:event_counts}

\noindent We are interested in modeling the statistical detection probability of particles emitted from a source. In our simulations we assume to have an ideal detector, meaning if a particle hits the detector the event is registered in the $i$-th pixel of a sky map with probability one.
Consequently, the conditional probability distribution for detector events can be expressed as:
\begin{equation}
\Pi(\mathrm{event}_i|\alpha_p,\delta_p, r_p) = W(\alpha_p,\delta_p, r_p) \, , 
\end{equation}
where $\alpha_p,\,\delta_p$ are the right ascension and declination of the $p$-th particle at the detector, and $r_p$ the distance passed by it at arrival time. The detector kernel $W(\alpha_p,\delta_p, r_p)$ is one for parameters that hit the detector and otherwise zero.
In fact, we are not interested in the conditional distribution but in the marginal distribution of events given as:
\begin{equation}
\Pi(\mathrm{event}_i) = \int \mathrm{d}\alpha_p \, \mathrm{d}\delta_p \, \mathrm{d}r_p \Pi(\mathrm{event}_i|\alpha_p,\delta_p, r_p) \, \Pi(\alpha_p,\delta_p, r_p) \, 
\end{equation}
where $\Pi(\alpha_p,\delta_p, r_p)$ is the distribution of arrival directions and distances to the detector as calculated by the CRPropa code. Note, that this distribution can be written as the marginal distribution over the initial conditions of the CRPropa simulation given by the emission angles $\alpha^{\mathrm{init}}_p,\delta^{\mathrm{init}}_p$ as:
\begin{equation}
\Pi(\alpha_p,\delta_p, r_p) = \int \mathrm{d}\alpha^{\mathrm{init}}_p \, \mathrm{d}\delta^{\mathrm{init}}_p \, \Pi(\alpha_p,\delta_p, r_p | \alpha^{\mathrm{init}}_p,\delta^{\mathrm{init}}_p ) \, \Pi(\alpha^{\mathrm{init}}_p,\delta^{\mathrm{init}}_p ) \, ,
\end{equation}
where $\Pi(\alpha^{\mathrm{init}}_p,\delta^{\mathrm{init}}_p )$ is the distribution of emission angles at the source. In general, for an isotropic source, emission angles are uniformly distributed over the full $4\pi$-geometry. Given these equations the distribution of detected events can be expressed as:
\begin{eqnarray}
\Pi(\mathrm{event}_i) &=& \int \mathrm{d}\alpha_p \, \mathrm{d}\delta_p \,  \mathrm{d}\alpha^{\mathrm{init}}_p \, \mathrm{d}\delta^{\mathrm{init}}_p \,  \mathrm{d}r_p \, \Pi(\mathrm{event}_i|\alpha_p,\delta_p, r_p) \, 
\Pi(\alpha_p,\delta_p, r_p | \alpha^{\mathrm{init}}_p,\delta^{\mathrm{init}}_p ) \, \Pi(\alpha^{\mathrm{init}}_p,\delta^{\mathrm{init}}_p ) \, \nonumber \\
&=& \int \mathrm{d}\alpha_p \, \mathrm{d}\delta_p \,  \mathrm{d}\alpha^{\mathrm{init}}_p \, \mathrm{d}\delta^{\mathrm{init}}_p \,  \mathrm{d}r_p \, W(\alpha_p,\delta_p, r_p) \,  \Pi(\alpha_p,\delta_p, r_p | \alpha^{\mathrm{init}}_p,\delta^{\mathrm{init}}_p ) \, \Pi(\alpha^{\mathrm{init}}_p,\delta^{\mathrm{init}}_p ) \, . 
\label{eq:event_integral}
\end{eqnarray}
In principle this is a statistical description of the CRPropa simulation framework: the CRPropa code simulates individual realizations of the joint distribution $\Pi(\alpha_p,\delta_p, r_p | \alpha^{\mathrm{init}}_p,\delta^{\mathrm{init}}_p ) \, \Pi(\alpha^{\mathrm{init}}_p,\delta^{\mathrm{init}}_p )$ via the following process:
\begin{itemize}
\item draw random emission angles at the source ($\alpha^{\mathrm{init}}_p,\delta^{\mathrm{init}}_p$)
\item follow particle trajectories to calculate the particle positions at arrival time ($\alpha_p,\delta_p, r_p$)
\end{itemize}
Note, that since it is possible to draw random realizations of the joint parameter set $\alpha_p,\delta_p, r_p,\alpha^{\mathrm{init}}_p$ and $\delta^{\mathrm{init}}_p$, it is possible to estimate the integral in equation \ref{eq:event_integral} via a Markov approximation as:
\begin{eqnarray}
\Pi(\mathrm{event}_i) &\approx& \frac{1}{N} \sum_p  W(\alpha_p,\delta_p, r_p)
\label{eq:event_est_stan}
\end{eqnarray}
where the index $p$ labels different random particle simulations performed with the CRPropa code. Also note, that the detection probability is proportional to the total intensity of observed particles at a given solid angle in the sky.
A particular issue arises from the fact that, when simulating particle propagation through cosmological volumes, it is very hard to find trajectories that will actually hit the rather small observer sphere. This is particularly owed to the fact that, due to physical considerations, it is reasonable to assume an isotropic emission of cosmic rays from the source. Consequently, a large fraction of randomly simulated particle trajectories is of no interest to predictions of observational properties since they miss the detector. This problem renders 3D and 4D cosmic-ray propagation simulations a numerically challenging problem.

The integral given in equation {\ref{eq:event_integral}} can also be solved in a slightly different fashion via an importance sampling approach:
\begin{eqnarray}
\Pi(\mathrm{event}_i) &=& \int \mathrm{d}\alpha_p \, \mathrm{d}\delta_p \,  \mathrm{d}\alpha^{\mathrm{init}}_p \, \mathrm{d}\delta^{\mathrm{init}}_p \,  \mathrm{d}r_p W(\alpha_p,\delta_p, r_p) \; \Pi(\alpha_p,\delta_p, r_p \,|\, \alpha^{\mathrm{init}}_p,\delta^{\mathrm{init}}_p ) \, \Pi(\alpha^{\mathrm{init}}_p,\delta^{\mathrm{init}}_p ) \, \nonumber \\
&=& \int \mathrm{d}\alpha_p \, \mathrm{d}\delta_p \,  \mathrm{d}\alpha^{\mathrm{init}}_p \, \mathrm{d}\delta^{\mathrm{init}}_p \,  \mathrm{d}r_p W(\alpha_p,\delta_p, r_p) \, \nonumber \\
& & \phantom{\int} \times\; \Pi(\alpha_p,\delta_p, r_p \,|\, \alpha^{\mathrm{init}}_p,\delta^{\mathrm{init}}_p ) \, \Pi'(\alpha^{\mathrm{init}}_p,\delta^{\mathrm{init}}_p ) \frac{\Pi(\alpha^{\mathrm{init}}_p,\delta^{\mathrm{init}}_p )}{\Pi'(\alpha^{\mathrm{init}}_p,\delta^{\mathrm{init}}_p )} \, 
\label{eq:event_integral_imp}
\end{eqnarray}
where in the last line we introduced a one and the probability distribution $\Pi'(\alpha^{\mathrm{init}}_p,\delta^{\mathrm{init}}_p )$ can now be freely chosen without changing the value of the integral. One can now draw realizations of the joint distribution $\Pi(\alpha_p,\delta_p, r_p | \alpha^{\mathrm{init}}_p,\delta^{\mathrm{init}}_p ) \, \Pi'(\alpha^{\mathrm{init}}_p,\delta^{\mathrm{init}}_p ) )$
with $\Pi'(\alpha^{\mathrm{init}}_p,\delta^{\mathrm{init}}_p )$ chosen such that it increases the probability of hitting the detector target.
The corresponding Markov approximation for the event distribution is then simply given by:
\begin{equation}
\Pi(\mathrm{event}_i) \;\approx\; \frac{1}{N} \sum_p  W(\alpha_p,\delta_p, r_p)\, \frac{\Pi(\alpha^{\mathrm{init}}_p,\delta^{\mathrm{init}}_p )}{\Pi'(\alpha^{\mathrm{init}}_p,\delta^{\mathrm{init}}_p )}\, \nonumber 
\;\;=\;\; \frac{1}{N} \sum_p  W(\alpha_p,\delta_p, r_p)\, \omega_p \, , 
\label{eq:event_est_gen}
\end{equation}
where the weight $\omega_p$ accounts for the modification of the statistical distribution of emission angles. Equation \ref{eq:event_est_gen}, therefore, is a generalization of the standard estimator presented in equation \ref{eq:event_est_stan}.

\section{A simple algorithm to target}

\noindent This work aims at providing optimal emission directions for sources in the CRPropa code, such that the yield of simulated particles trajectories hitting a distant detector is maximized. Because particles do not necessarily travel in straight lines, but their trajectories may be bent due to magnetic fields or other kinds of interactions, the optimal emission direction at the source does not necessarily point directly to the source. Moreover, random magnetic fields or scattering events can broaden the emitted particle distribution, such that the shape of the target looks blurred from the perspective of the source. For an arbitrary source in a general CRPropa simulation, the exact optimal direction and the width of the emission direction are not known and need to be identified on the fly during runtime.

We describe the optimal emission probability distribution for arbitrary distant sources by the von Mises-Fischer (vMF) distribution
\begin{equation}
\label{eq:vMF_hit_prob}
\Pi(x|\mathrm{hit}) = \left\{
    \begin{array}{ll}
      \displaystyle \frac{1}{4\pi}, & \qquad \kappa=0 \\
      \displaystyle \frac{\kappa}{2\pi \left(1-\mathrm{e}^{-2\kappa} \right)} \mathrm{e}^{\kappa \left( \vec{\mu}^T\vec{x}-1\right)}, & \qquad \kappa>0
    \end{array}\right. \, ,
\end{equation}
where $\vec{\mu}$ is the Cartesian unit vector corresponding to the preferred direction of emission,  $\vec{x}$ is a unit vector of a random direction on the 2-sphere, and  $\kappa$ controls the width of the distributions of random emission directions around the preferred direction. In particular we choose $\kappa$ simply from requiring that in the case of geometrical optics a fraction $P$ of emitted particles
would hit the observer sphere. More specifically, we need to solve the equation
\begin{equation}
P \;=\; \int_a^1 \mathrm{d}y\;\;\frac{\kappa}{2\pi \left(1-\mathrm{e}^{-2\kappa} \right)}\;\mathrm{e}^{\kappa \left( y -1 \right)} \;\;=\;\;  \frac{1-\mathrm{e}^{\kappa \left(a-1 \right)}}{1-\mathrm{e}^{-2\kappa}}\, .
\end{equation}
If we assume $\kappa \gg 1 $ we can find the approximation
\begin{equation}
\label{eq:estimate_kappa}
\kappa = \frac{\mathrm{ln}(1-P)}{a-1}\, ,
\end{equation}
where $a=\mathrm{arctan}(s/D)$ is the apparent detector size, with $s$ the radius of the observer sphere and $D$ the distance between the source and the detector. By choosing the hit probability $P$ one can now adjust the amount of simulated particles that are expected to arrive at the detector. When, e.g., the particle trajectories are getting close to the diffusion regime, the naive geometric targeting will be sub-optimal, hence one should choose $P$ to have a suitable ratio of exploitation and exploration. In practice, setting $P$ is a matter of choice and of no importance to the validity of the algorithm. In the large sample limit the algorithm will converge to the correct result as the importance weights $\omega_n$ ensure that the algorithm simulates the correct emission statistics asymptotically. Typical values may range from $P=0.1$ if we want to be conservative and be sure not to miss any unexpected paths to the target, to $P=0.9$ when we are expect to be close to the case of geometrical optics. 

To find the optimal emission direction and the width of the distribution, i.e., the parameters $\vec{\mu}$ and $\kappa$, we run the simulation in terms of several epochs $N_{\mathrm{epoch}}$, where each epoch simulates a batch of $N_{\mathrm{batch}}$ simulation particles. The directions of emission at the source have been 
drawn from a vMF distribution, using algorithms readily discussed in the literature~\cite{Ulrich:1984:CGD,tROB04a}. We record these emission directions and follow particle trajectories through the simulation. At the end of every batch of simulations we determine new optimal parameters for the vMF and run a new batch with updated parameters. To learn the target distribution the algorithm will only rely on those emission directions $\vec{x}_n$ whose trajectories will end at the detector surface. Since these particles have been emitted according to a vMF distribution, while the true physical source would emit uniformly over $4\pi$, we have to estimate their importance weights as
\begin{equation}
\omega_n \;=\; \frac{1}{4\pi} \, \frac{1}{\Pi(x\,|\,\mathrm{hit})} 
\;\;=\;\; \frac{ \left(1-\mathrm{e}^{-2\kappa} \right)}{2\,\kappa} \mathrm{e}^{-\kappa \left( \vec{\mu}^T\vec{x}-1\right)} \, . 
\end{equation}
Given successful emission directions $\vec{x}_n$ and the corresponding importance weights $\omega_n$ we may now estimate the preferred emission direction as
\begin{equation}
\vec{\mu} = \frac{\sum_{n=0}^{N_{\mathrm{batch}}} \vec{x}_n \, \omega_i  }{ \left|\sum_{n=0}^{N_{\mathrm{batch}}} \vec{x}_n \, \omega_n\right|} \, .
\end{equation}
the detector size $a$ by
\begin{equation}
a = \frac{\sum_{n=0}^{N_{\mathrm{batch}}} \left(\vec{\mu}^T \vec{x}_n \right)^2 \, \omega_i  }{\sum_{n=0}^{N_{\mathrm{batch}}}  \omega_n}\, 
\end{equation}
and the corresponding optimal parameter $\kappa$ of the vMF distribution is then found from eq.~\ref{eq:estimate_kappa}.

\section{Speedup test}

\begin{figure*}[tb]
\centering
 \centering
  \includegraphics[width=.49\linewidth]{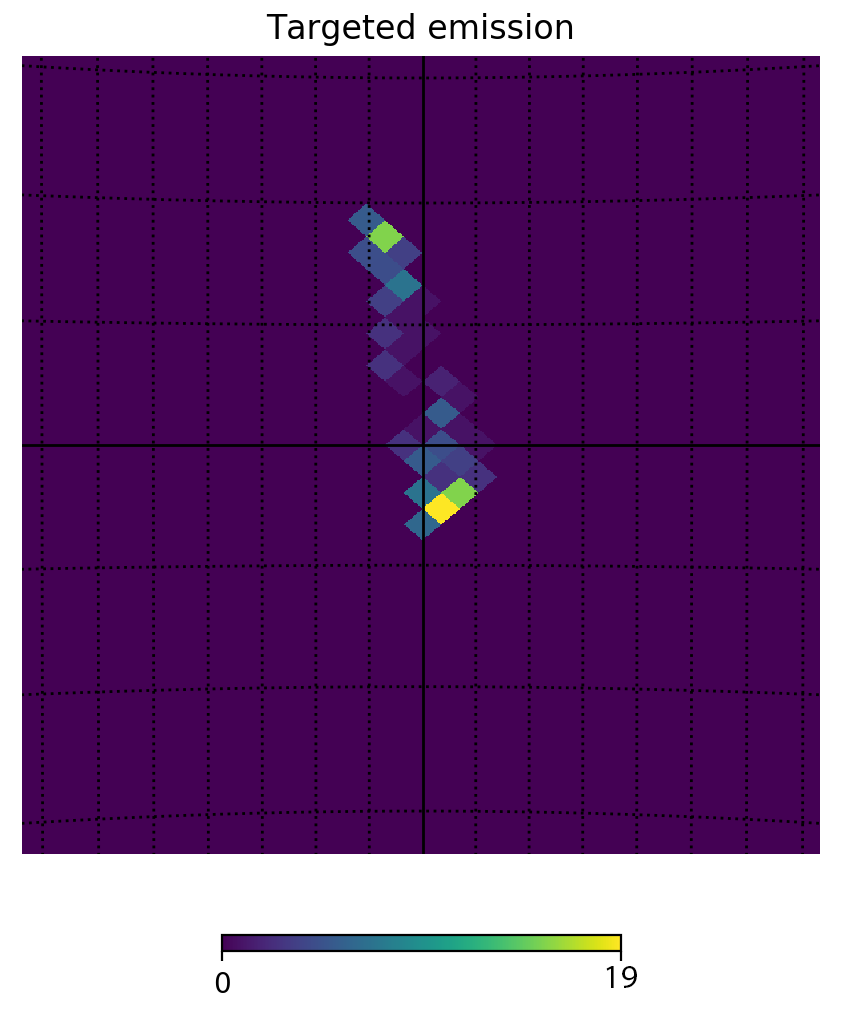}
  \includegraphics[width=.49\linewidth]{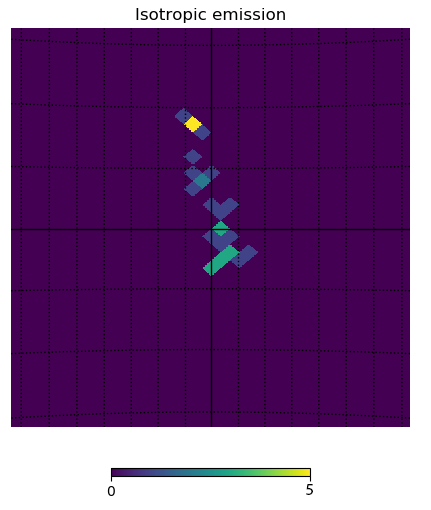}
  \caption{Comparison of sky maps for the reference simulation: one spherical observer with a radius of $R_{\mathrm{obs}} = 0.1$ Mpc, one source at a distance of $D = 10$ Mpc, emitting 1M protons with an energy of $E = 10$ EeV, a Kolmogorov-type turbulent magnetic field with $B_{\mathrm{RMS}} = 1$ nG,  a hit probability of $P = 0.1$ and no interactions included. Left: the map obtained using targeted emission. Right: the map obtained using isotropic emission. In both cases the same number of particles were emitted from the source. The total number of hits is $\sim100$ times larger in the left figure compared with the right figure.}\label{Fig:Skymaps}
\end{figure*}

\begin{figure*}[tb]
\centering
 \centering
  \includegraphics[width=0.82\linewidth]{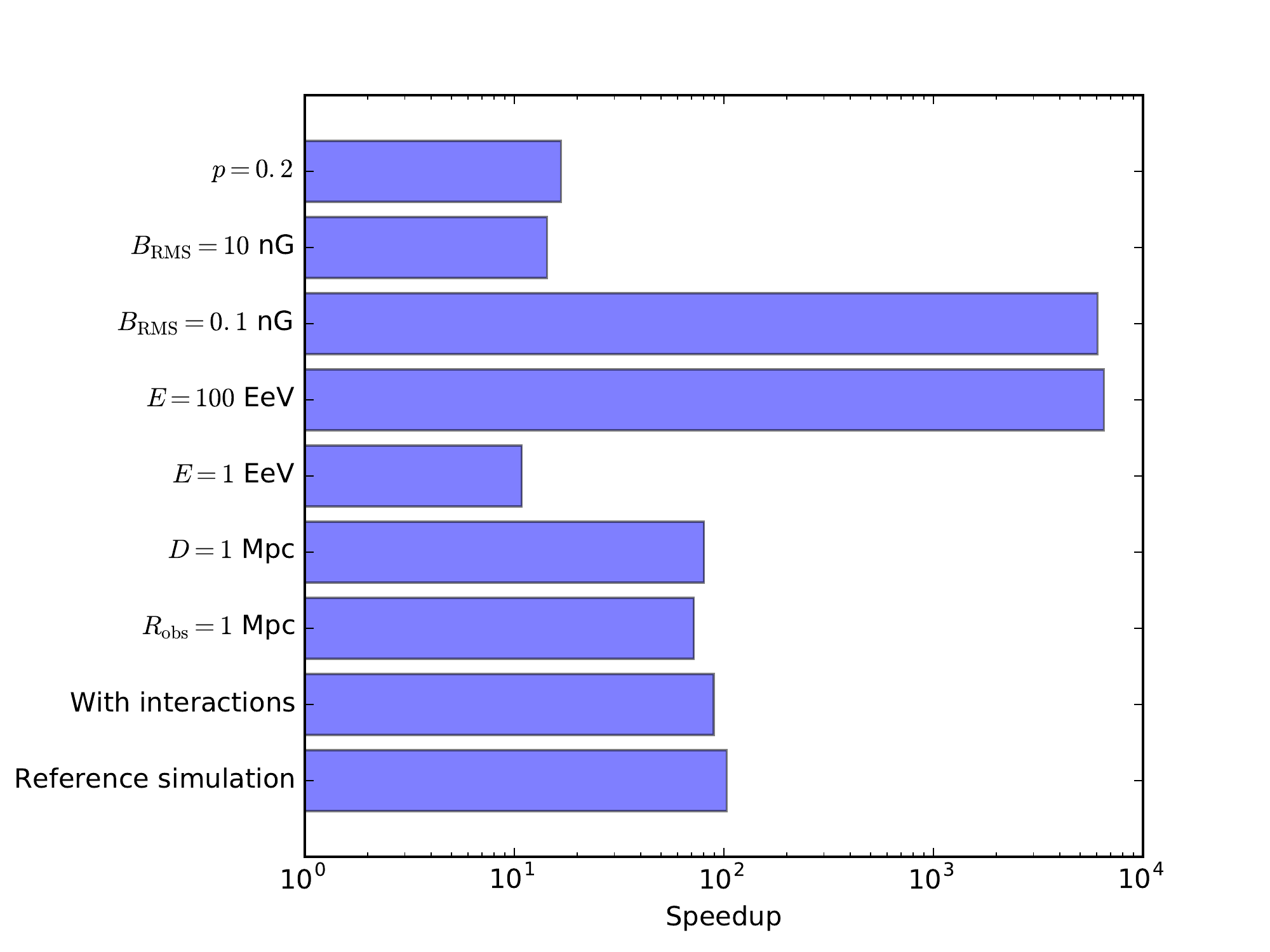}
  \caption{Test of the speedup of CRPropa using directed emission. Reference simulation: one spherical observer with a radius of $R_{\mathrm{obs}} = 0.1$ Mpc, one source at a distance of $D = 10$ Mpc, emitting 1M protons with an energy of $E = 10$ EeV, a Kolmogorov-type turbulent magnetic field with $B_{\mathrm{RMS}} = 1$ nG,  a hit probability of $p = 0.1$ and no interactions included. For the other scenarios one parameter of the reference simulation is changed while all other are kept the same.}\label{Fig:Speedup}
\end{figure*}

\noindent To test the speedup that targeted emission only (without the learning method described above for determining the optimal emission settings) can provide, we run CRPropa simulations with exactly the same settings, only changing between targeted emission and isotropic emission. The speedup is then defined as 
\begin{equation}
\mathrm{S} = \frac{t_{\mathrm{iso}}/N_{\mathrm{iso}}}{t_{\mathrm{tar}}/N_{\mathrm{tar}}}
\end{equation}
with $t_{\mathrm{iso}}$ ($t_{\mathrm{tar}}$) and $N_{\mathrm{iso}}$ ($N_{\mathrm{tar}}$) the time the simulation took and the number of hits at the observer for isotropic (targeted) emission. For this test we run a reference simulation with one spherical observer with a radius of $R_{\mathrm{obs}} = 0.1$~Mpc and one source at a distance of $D = 10$~Mpc emitting protons with an energy of $E = 10$~EeV in a Kolmogorov-type turbulent magnetic field with $B_{\mathrm{RMS}} = 1$~nG and a maximum correlation length of 1~Mpc. The hit probability $P$ has not been optimised but has arbitrarily been set to $P=0.1$. For this reference simulation all interactions have been switched off. In both the isotropic and targeted emission cases 1M particles have been emitted from the source. The speedup found in this case is $S \approx 103$, already a huge computational improvement without any optimization of $P$ or other simulation settings. A comparison between the two sky maps that were obtained for this reference simulation, one for targeted emission and one for isotropic emission with the same number of particles emitted from the source, is given in Fig.~\ref{Fig:Skymaps}.

To see how the simulation parameters $R_{\mathrm{obs}}$, $D$, $E$, $B_{\mathrm{RMS}}$ and $p$ and the inclusion of interactions with photon backgrounds influence the speedup, we change one parameter of the reference simulation at a time and recalculate the speedup. The results of this procedure are given in Fig.~\ref{Fig:Speedup}. This shows that, for the scenarios tested here, the speedup can vary between $S \approx 10$ and $S \approx 6500$ depending on the simulation setup. This should be considered as a minimal speedup as no auto-tuning of $p$ or the emission direction has been included.

\section{Further development of the method and implementation in CRPropa}

\noindent In our approach so far the desired hit probability $P$ was a matter of choice, and the vMF parameter $\kappa$ and $\vec{\mu}$ were optimized only with respect to ensuring that the prior choice of $P$ matches the posterior hit probability in the given simulation setup. Certainly there is an optimal choice of $P$ for every setup, so it would be interesting to find this value by simulations, which can be done by Bayesian sampling. In practice it means to maximize the logarithmic posterior distribution given as
\begin{equation}
\label{eq:log_meta_post}
\ln\left( \phantom{\big(}\!\!\!\Pi\left(\kappa,\vec{\mu},P\,|\,\{\vec{x}\}_{\rm hit},\{\vec{x}\}_{\rm nohit}\right)\phantom{\big)}\!\!\!\!\!\right) \;\;=\;\; 
\sum_{p=0}^{N_{\rm hit}} \ln\left( \Pi(\vec{x}_p\,|\,{\rm hit}) \right) + 
\sum_{q=0}^{N_{\rm nohit}} \ln\left(\Pi(\vec{x}_q\,|\,{\rm nohit})\right)\quad, 
\end{equation}
where $\Pi(\vec{x}_p\,|\,{\rm hit})$ is the vMF invoked for the parameters $P$, $\kappa$ and $\vec{\mu}$ chosen for the batch, 
\begin{equation}
\Pi(\vec{x}\,|\,{\rm nohit}) 
\;=\; \frac{1}{1-P}\;\left(1- \frac{P\kappa}{2\pi \left(1-\mathrm{e}^{-2\kappa} \right)} \mathrm{e}^{\kappa \left( \vec{\mu}^T\vec{x}-1\right)}\right)\;, 
\end{equation}
and $p$ and $q$ label particles that hit or miss the target, respectively. Note that this way we use information from all particles simulated in the batch, if we only had particles hitting the target the task of finding improved meta parameters would reduce to a standard vMF regression with $P$ as a free parameter. Optimal maximum a posteriori values for $P$ and the corresponding parameters $\kappa$ and $\vec{\mu}$ are then found by employing a standard numerical optimizer to maximize equation {\ref{eq:log_meta_post}}. 

The functionality described here will be made publicly available in the near future as part of the main repository of CRPropa.\footnote{
\href{https://crpropa.desy.de/}{crpropa.desy.de}}
The targeting algorithm using the vMF distribution will be implemented as an extension of the {\sc Source} module of CRPropa. The learning routine to obtain the optimal parameters of the vMF distribution will be added to the main repository as a plugin to the code structure. Instructions on how to use the targeting algorithm, together with the learning routine, will be made available on the CRPropa website as part of a dedicated example page. More information on the method and its capabilities will be given in an upcoming journal publication (Jasche, van Vliet \& Rachen, in preparation).

\begin{footnotesize}

\subsection*{Acknowledgements}

\noindent This research was supported in part by the DFG cluster of excellence \textit{Origin and Structure of the Universe}.\footnote{
\href{http://www.universe-cluster.de/}{www.universe-cluster.de}} 
AvV acknowledges financial support from the NWO Astroparticle Physics grant WARP and the European Research Council (ERC) under the European Union's Horizon 2020 research and innovation programme (Grant No. 646623).

\end{footnotesize}

\end{document}